# Confinement Effect on Thermopower of Electrolytes

Xin Qian[1]*, Te-Huan Liu[1], and Ronggui Yang[1,2]*

[1]School of Energy and Power Engineering,

Huazhong University of Science and Technology, Wuhan 430074, China

[2]State Key Labaratory of Coal Combustion,

Huazhong University of Science and Technology, Wuhan 430074, China

Corresponding Email:

xinqian21@hust.edu.cn

ronggui@hust.edu.cn

## Abstract

Ionic Seebeck effect of electrolytes has shown promising applications in harvesting energy from low-grade waste-heat sources with small temperature difference from the environment, which can power sensors and Internet-of-Things devices. Recent experiments have demonstrated giant thermopower (~ 10 mV/K) of electrolytes under confinement due to the overlapping of electric double layer (EDL). Nonetheless, there has been no consensus on the theory of the ionic Seebeck effect, especially whether the thermopower depends on ionic diffusivities, imposing confusion on the theoretical interpretation of experimental discovery on giant thermopower of confined electrolytes. This article presents a linear perturbative solution of Poisson-Nernst-Planck (PNP) equations to describe the ionic Seebeck effect of confined liquid electrolytes. We provide both analytical and numerical solutions to the PNP equations for closed systems and open systems connected to reservoirs of electrolytes. The analytical solution captured the confinement effect both along and perpendicular to the temperature gradient, and showed excellent agreement with numerically solved PNP equations for a wide range of EDL potentials, channel widths, and lengths. Finally, we show that for polyelectrolytes with largely mismatched diffusivities, thermopower can only be enhanced for the closed system through confinement perpendicular to the temperature gradient.

## I. Introduction

The ionic Seebeck effect, also known as the Soret effect, [1, 2] is analogous to the thermoelectric effect in semiconductors that generates an electromotive force due to the thermodiffusion of charge carriers when a temperature gradient is applied. Equivalent to the electrons or holes in semiconductors, the charge carriers responsible for the ionic Seebeck effect are mobile cations and anions of the electrolyte. Early studies in ionic Seebeck effect focused on bulk electrolytes, with small thermopowers of only a few tens of μV/K,[3] which is responsible for the thermophoresis phenomena of colloidal solution.[4, 5] In the past few years, ionic liquids, gel and liquid electrolytes inside polymer matrices have shown significantly improved thermopower, on the order of 1~24 mV/K (nearly $10^2 \, k_B/e$ with $k_B$ the Boltzmann constant and $e$ the elementary charge).[6-11] With such high thermopower, there is great potential for ionic systems to harvest energy from low-grade waste-heat sources with small temperature difference from the environment. For example, Han and Qian *et al.*[7] demonstrated that a high voltage of 2V can be generated with only 25 ionic gelatin modules in series for harvesting body heat, which can supply power to the Internet-of-Things sensors.

However, there is no consensus on the theory of the ionic Seebeck effect. One most commonly used formalism for ionic thermopower of a symmetrical electrolyte is $S = \frac{k_B}{ze}(\alpha_+ - \alpha_-)$, where $z$ is the valence of ions, $k_B$ and $e$ the Boltzmann constant and elementary charge, $\alpha_\pm$ the dimensionless Soret coefficients correlating the temperature gradient and the thermally induced concentration gradient $\nabla(\ln n_\pm) = -2\alpha_\pm \nabla(\ln T)$.[12] Expressions of the ionic thermopower is obtained setting ionic flux as zero: $J_\pm = -D_\pm \left( \nabla n_\pm \pm \frac{zen_\pm}{k_B T} \nabla \phi + \frac{2\alpha_\pm n_\pm}{T} \nabla T \right) = 0$, with $D, n$ the diffusivity and the concentration, $\phi$ the electric potential, while assuming local charge neutrality $\nabla(n_+ - n_-) = 0$.[9, 13] Such an approximation of local charge neutrality is, however, intrinsically contradictory to the Poisson equation of the electrostatic theory, which was also pointed out by Chikina *et al.*[14] If local charge neutrality held everywhere inside the electrolyte, there would not be any non-homogeneity of the electric potential and no thermal voltage could be measured. This is more prominent for electrolytes confined between boundaries or inside porous media with characteristic lengths comparable to the Debye length of the electrolyte (usually a few nanometers), *i.e.,* the assumption of local charge neutrality breaks down.[15] Other theories that the ionic thermopower depends on the ionic mobilities or transference numbers of the ionic species. Such derivations involve the zero-current condition $z(J_+ - J_-) = 0$ and the similar assumption of homogeneous concentration profile $\nabla n_\pm = 0$, which are incompatible to the electrostatic theory. [7, 16-18] Recently, Sehnem and Janssen pointed out that ionic thermopower depends on ionic mobilities only in the early response after applying a temperature gradient, and the steady state thermopower becomes independent of ionic mobilities.[19] Würger pointed out that theoretical expressions for the thermopower depend on whether the electrolyte is allowed to exchange currents with the reservoirs.[20] For a closed system, the temperature gradient drives the ionic species to migrate and accumulate on the cold electrode. At the steady state, the mismatch of the Soret coefficient would result in different concentration gradients of cations and anions, hence a local charge density profile will be developed. In this closed system, the thermopower is solely dependent on the Soret coefficients. For an open system, the ionic current is allowed to exchange between the electrolyte and the reservoir, resulting in mobility dependence of the thermopower.

There yet exist quantitative models for the confinement effect on ionic thermopower for both open and closed systems. Previous models for the thermoelectric effect of confined electrolytes focused on infinitely long open channels,[21] indicating that the mismatch of ionic mobilities induced by confinement would increase the thermopower. However, the experimental measurements for confined electrolytes are performed with ionic-insulating electrodes,[22] whose thermopower are independent of ionic mobilities.[19, 20] It is questionable whether the theory for open systems[21] can be used to interpret the experimental results because most of such measurements on thermopower are performed in closed systems if there are no redox-active species.

In this article, theoretical expressions for ionic thermopower are derived by solving the Poisson-Nernst-Planck (PNP) equations with first order perturbation. First, we rigorously solve the PNP equations in the one-dimensional (1D) limit, showing that the thermopower depends on whether the system is closed (Figure 1a) or open, *i.e.* connected to solution reservoirs, as shown in Figure 1b. The 1D solution also uncovered the size effect of the thermopower in the axial direction along the temperature gradient. Then a set of two-dimensional (2D) partial differential equations (PDEs) is derived, capturing the confinement effect in the lateral direction perpendicular to the temperature gradient for both closed (Figure 1c) and open systems (Figure 1d). The detailed derivations will be presented in Section II. In Section III, we perform numerical validation for the analytical results of ionic thermopower in a wide range of channel widths, lengths, and EDL potentials, which showed excellent agreement with the analytical model. We also provide insights into the thermopower of confined polyelectrolytes, with extremely mismatched ionic mobilities. We show that in such ionic electrolytes with one species almost immobile, confinement can improve the thermopower only for a closed system. This work provides a theoretical benchmark for interpreting the ionic thermopower for liquid electrolytes confined in nanochannels.

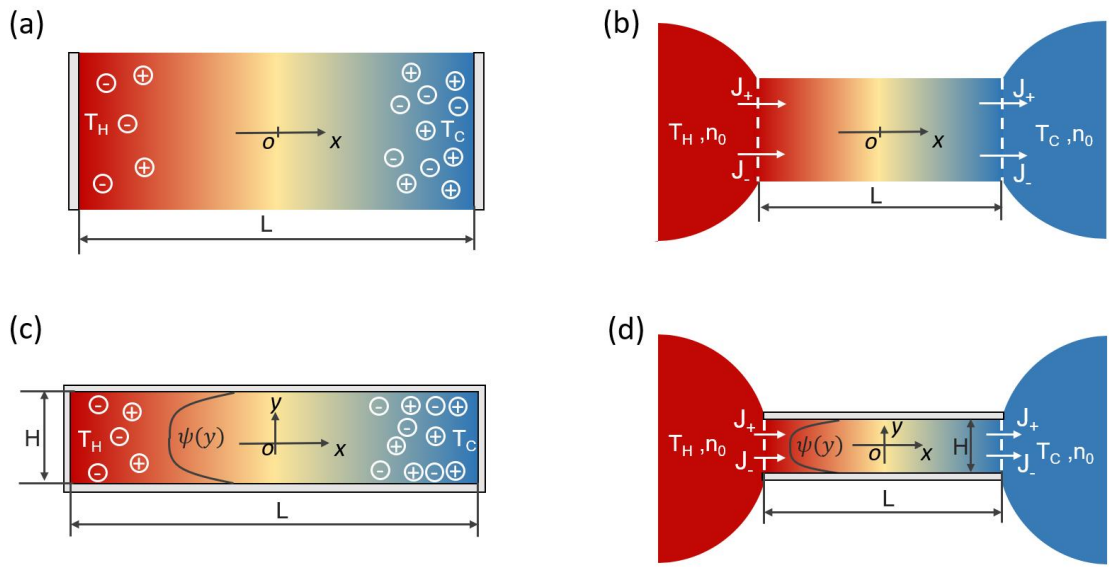

Figure 1. Schematic of (a) a 1D closed system and (b) a 1D open system connected to two reservoirs at different temperatures $T_H$ and $T_C$ but with the same concentration $n_0$. (c) Schematic of a 2D closed system. (d) Schematic of a 2D system connected to two reservoirs. The boundaries indicated by light gray areas are ionic-insulating, such that no ionic current is allowed across the interfaces. Confinement in the lateral direction perpendicular to the temperature gradient results in an electric double layer potential $\psi(y)$.

## II. Theoretical Model for Ionic Seebeck Effect in Confined Electrolyte

This section presents the theoretical derivation for ionic thermopower in electrolytes under 1D and 2D confinement. To begin with, governing equations are derived based on the conservation laws and the Onsager transport theory in part A, then the solutions to the PNP equations in 1D and 2D limits are presented in parts B and C, leading to analytical expressions of ionic thermopower for both open and closed systems confined by ionic insulating boundaries.

### A. Governing equations of coupled thermal-ionic transport

To begin with the derivation, we write the Onsager transport equations for coupled thermal-ionic transport, in which the ionic flux $\boldsymbol{J}_i$ and heat flux $\boldsymbol{J}_Q$ are expressed as linear combinations of the thermodynamic forces,[23, 24]

$$\boldsymbol{J}_i = L_{ii}\left(-\frac{\nabla \bar{\mu}_i}{T}\right) + L_{iQ}\nabla\left(\frac{1}{T}\right)$$

$$\boldsymbol{J}_Q = \sum_i L_{Qi}\left(-\frac{\nabla \bar{\mu}_i}{T}\right) + L_{QQ}\nabla\left(\frac{1}{T}\right) \quad (1),$$

where $T$ denotes the temperature $\bar{\mu}_i$ is the electrochemical potential, defined as $\bar{\mu}_i = \mu_i + z_i e\phi$, with $\mu_i$ the chemical potential, $z_i$ the valance charge of species $i$ and $\phi$ the electrostatic potential. The coefficients $L_{ii}$, $L_{iQ}$, $L_{Qi}$ and $L_{QQ}$ are the linear transport coefficients. Onsager reciprocity demands $L_{iQ} = L_{Qi}$ to satisfy the time-reversal symmetry of microscopic kinetics. Since the chemical potential is a functional of concentration and temperature profiles: $\mu_i = \mu[n_i(\boldsymbol{r}), T(\boldsymbol{r})]$, we can expand the gradient of chemical potential as $\nabla \mu_i = \left(\frac{\partial \mu_i}{\partial n_i}\right)_T \nabla n_i + \left(\frac{\partial \mu_i}{\partial T}\right)_{n_i} \nabla T$. With $\left(\frac{\partial \mu_i}{\partial n_i}\right)_T = \frac{k_B T}{n_i}$ and $\left(\frac{\partial \mu_i}{\partial T}\right)_{n_i} = -s_i$ where $s_i$ is the partial entropy of species $i$, the ionic flux can be rewritten as,

$$J_i = L_{ii}\left(-\frac{k_B}{n_i}\nabla n_i - \frac{z_i e}{T}\nabla\phi + \frac{s_i}{T}\nabla T\right) - \frac{L_{iQ}}{T^2}\nabla T \qquad (2).$$

Defining the transported heat $Q_i^*$ as:[3]

$$Q_i^* = L_{Qi}/L_{ii}, \qquad (3)$$

the ionic flux can be written as:

$$J_i = -\frac{L_{ii}k_B}{n_i}\left(\nabla n_i + \frac{z_i e}{k_B T}\nabla\phi + \frac{Q_i^* - s_i T}{k_B T^2}n_i\nabla T\right) \qquad (4)$$

Further, the dimensionless Soret coefficient is defined as,

$$\alpha_i = \frac{Q_i^* - s_i T}{2k_B T} = \frac{\hat{s}_i}{2k_B} \qquad (5)$$

where $\hat{s}_i$ is the Eastman entropy of transfer, defined as $\hat{s}_i = Q_i^*/T - s_i$. It is important to note that this definition of the dimensionless Soret coefficient and the Eastman entropy of transfer is different from the original definition by Eastman,[25] but is similar to the approach by Agar[3] and Huang *et al.*[17] based on Onsager transport theory. From Eq. (4), it is clear that the Soret effect is a combination of both nonequilibrium transport properties and equilibrium thermodynamics such as solvation effects and electric double layers.

With the above definition of the Soret coefficient, the constitutive relation for ionic flux is obtained as,

$$J_i = -D_i\left(\nabla n_i + \frac{z_i e n_i}{k_B T}\nabla\phi + \frac{2n_i\alpha_i}{T}\nabla T\right), \qquad (6),$$

where $D_i = L_{ii}k_B/n_i$ is the diffusivity of ionic species $i$. In a closed system with zero ionic flux, the dimensionless Soret coefficient $\alpha_i$ correlates the temperature gradients with the concentration gradients at the limit of low electric field $\nabla\phi \approx 0$ as:

$$\frac{\nabla n_i}{n_i} + 2\alpha_i \frac{\nabla T}{T} = 0 \tag{7}$$

Similarly, we can obtain the relation for heat flux:

$$J_Q = \sum_i Q_i^* J_i - k\nabla T \tag{8}$$

where $k = (L_{QQ} - \sum_i L_{Qi}L_{iQ}/L_{ii})/T^2$ is the thermal conductivity. From eq.(7) we can see that $Q_i^*$ is the amount of heat carried along with the ionic flux of species $i$. Eq (5) and (7) are the constitutive equations for coupled thermal-ionic transport. However, the second and third terms in Eq. (5) would result in nonlinear PDEs for the conservation laws, which imposes great challenges for both analytical and numerical solutions.

To proceed with the modeling of coupled thermal-ionic transport, we strive to linearize the constitutive equations by taking the first-order perturbation approach. The field variables $n$, $\phi$, and $T$ can be separated into the equilibrium homogenous part (denoted with subscript 0) and the spatially varying nonequilibrium part (denoted with subscript 1),

$$n_i(\mathbf{r}) = n_{i0} + n_{i1}(\mathbf{r})$$
$$\phi(\mathbf{r}) = \phi_0 + \phi_1(\mathbf{r}) \tag{9}$$
$$T(\mathbf{r}) = T_0 + T_1(\mathbf{r})$$

The coupled thermal-ionic transport is assumed to be near equilibrium, such that the perturbation is small and the ionic flux can be linearized,

$$\mathbf{J}_i = -D_i\left(\nabla n_{i1} + \frac{z_i e n_{i0}}{k_B T_0}\nabla\phi_1 + \frac{2\alpha_i n_{i0}}{T_0}\nabla T_1\right) \tag{10}$$

Finally, the governing equations for the coupled thermal-ionic transport are obtained by combining the conservation laws of species and energy with Poisson's equation of electrostatics:

$$\frac{\partial n_{i1}}{\partial t} + \nabla \cdot \boldsymbol{J}_i = 0 \tag{11}$$

$$\rho c_p \frac{\partial T_1}{\partial t} - k\nabla^2 T_1 = \sum_i Q_i^* \frac{\partial n_{i1}}{\partial t} \tag{12}$$

$$\nabla^2 \phi_1 = -\epsilon^{-1} \sum_i z_i e n_{i1} \tag{13}$$

where $\epsilon$ is the dielectric constant of the electrolyte. Interestingly, the transported heat $Q_i^*$ couples the temperature field with the concentration profile, with the transient variance of concentration acting like a source term in the heat diffusion equation.[26] At the steady-state, the heat diffusion equation becomes homogenous and is decoupled from the concentration field. In this paper, we seek steady-state solutions to obtain the expressions for the ionic thermopower confined by boundaries.

### B. Confinement Effect on Thermopower in 1D Limit

Before directly solving the full two-dimensional PNP equations, it is helpful to outline the logistics of obtaining the thermopower in 1D transport limit as shown in Figure 1a-b, where a temperature gradient is imposed on the length scale $L$. We rigorously show that these two cases would result in different expressions of thermopower. For the closed system, the thermopower is solely determined by the Soret coefficient, while the thermopower for an open system is dependent on ionic diffusivities. The result uncovered the size dependence along the direction of temperature gradient. When the distance between the two electrodes is much larger than the Debye length $\kappa^{-1}$, the thermopower converged to the bulk limit, agreeing well with results derived by Würger.[20]

In the 1D limit, the conservation law of species at steady state $\nabla \cdot \boldsymbol{J}_i = 0$ is written as:

$$\frac{d^2 n_{i1}}{dx^2} + \frac{z_i e n_{i0}}{k_B T_0}\frac{d^2\phi_1}{dx^2} + \frac{2\alpha_i n_{i0}}{T_0}\frac{d^2 T_1}{dx^2} = 0 \tag{14}$$

The energy conservation $\nabla \cdot \boldsymbol{J}_Q = 0$ would simply result in,

$$\frac{d^2 T_1}{dx^2} = 0 \tag{15}$$

If we apply a fixed $T_H$ at the left boundary and $T_C$ at the right boundary, the temperature profile is simply linear, $T_1(x) = -\frac{\Delta T}{L}x$, with $\Delta T = T_H - T_C$. The electric potential is determined by the Poisson equation:

$$\frac{d^2 \phi_1}{dx^2} = -\epsilon^{-1}\sum_i z_i e n_{i1} \tag{16}$$

Without losing any physical insights, the electrolyte is assumed symmetrical for the simplicity of the solution, such that $z_+ = -z_- = z$, $n_{+0} = n_{-0} = n_0$. To capture the effect of mismatched ionic diffusivity or mobilities, this work does not assume equal diffusivity as most others have done.[21, 27] We then arrive at the following set of equations:

$$\frac{d^2 n_{+1}}{dx^2} + \frac{z e n_0}{k_B T_0}\frac{d^2\phi_1}{dx^2} = 0 \tag{17}$$

$$\frac{d^2 n_{-1}}{dx^2} - \frac{z e n_0}{k_B T_0}\frac{d^2\phi_1}{dx^2} = 0 \tag{18}$$

$$\frac{d^2\phi_1}{dx^2} = -\frac{ze}{\epsilon}(n_{+1} - n_{-1}) \tag{19}$$

Eqs. (17-19) will be solved to show how the boundary conditions would affect the final expressions of thermopower. By replacing $n_{\pm 1}$ with $n_d = \frac{1}{2}(n_{+1} - n_{-1})$ and $n_m = \frac{1}{2}(n_{+1} + n_{-1})$, together with Eq. (19), Eqs. (17-18) can then be simplified as:

$$\frac{d^2 n_m}{dx^2} = 0 \tag{20}$$

$$\frac{d^2 n_d}{dx^2} - \kappa^2 n_d = 0, \tag{21}$$

where $\kappa^2 = 2z^2 e^2 n_0 / \epsilon k_B T_0$, and $\kappa^{-1}$ is known as the Debye length beyond which electrostatic interaction is screened by mobile ionic charges. The general solution of Eq. (21) is the linear combination of $\sinh(\kappa x)$ and $\cosh(\kappa x)$. By requiring charge neutrality, $\int_{-L/2}^{L/2} n_d dx = 0$, only the odd hyperbolic sine function survives, hence $n_d$ is proportional to $\sinh(\kappa x)$, and the proportionality coefficient is to be determined by the boundary conditions.

In a closed system as shown in Figure 1a, ionic species are not allowed to cross the boundaries, hence the fluxes are zero. Comparing with electrolytes, the metal electrode is usually orders of magnitudes more conductive, and the field at metal electrode surfaces is negligibly small. The boundary conditions are:

$$J_i\left(x = \pm \frac{L}{2}\right) = 0$$

$$\frac{\partial \phi_1}{\partial x}\left(x = \pm \frac{L}{2}\right) = 0 \tag{22}$$

Expanding the ionic fluxes into the gradients of concentration, temperature, and electric field, we can obtain boundary conditions for $n_d$ and $n_m$:

$$\frac{\partial n_d}{\partial x}\left(x = \pm\frac{L}{2}\right) = \frac{(\alpha_+ - \alpha_-)n_0}{T_0}\frac{\Delta T}{L} \quad (23)$$

$$\frac{\partial n_m}{\partial x}\left(x = \pm\frac{L}{2}\right) = \frac{(\alpha_+ + \alpha_-)n_0}{T_0}\frac{\Delta T}{L} \quad (24)$$

Therefore, concentration profiles can be solved as:

$$n_d(x) = n_0 \frac{(\alpha_+ - \alpha_-)\Delta T}{T_0} \frac{\sinh(\kappa x)}{\kappa L \cosh\left(\frac{\kappa L}{2}\right)} \quad (25)$$

$$n_m(x) = \frac{(\alpha_+ + \alpha_-)n_0}{T_0}\frac{\Delta T}{L}x \quad (26)$$

By integrating the Poisson equation with the boundary condition $\frac{\partial \phi}{\partial x}\left(x = \pm\frac{L}{2}\right) = 0$, the electric field $E$ is obtained:

$$-E = \phi'(x) = \frac{(\alpha_+ - \alpha_-)\Delta T k_B}{zeL}\left(1 - \frac{\cosh(\kappa x)}{\cosh\left(\frac{\kappa L}{2}\right)}\right) \quad (27)$$

Another integration of $\phi'$, we can obtain the potential difference:

$$\phi_C - \phi_H = \frac{(\alpha_+ - \alpha_-)(T_H - T_C)k_B}{ze}\left(1 - \frac{\tanh\xi}{\xi}\right) \quad (28)$$

where $\xi = \kappa L/2$, $\phi_C$ and $\phi_H$ are the electric potential of the cold electrode surface at $T_C$ and the hot electrode surface at $T_H$. Finally, the Seebeck coefficient can be derived:

$$S_{1D}(\xi) = -\frac{\phi_H - \phi_C}{T_H - T_C} = \frac{k_B}{ze}(\alpha_+ - \alpha_-)\left(1 - \frac{\tanh\xi}{\xi}\right) \quad (29)$$

In closed systems, the Seebeck coefficient is determined by the mismatch of the Soret coefficient between cations and anions, and the thermopower is a length-dependent quantity. When the distance between the two electrodes is comparable to the Debye length $\kappa^{-1}$, the thermopower is suppressed. When the distance between two electrodes is much longer than the Debye length $\kappa^{-1}$, such that $\kappa L \gg 1$, the bulk Seebeck coefficient will converge to the bulk limit:

$$S_{bulk} = \frac{k_B}{ze}(\alpha_+ - \alpha_-) \tag{30}$$

In open systems connected to reservoirs as shown in Figure 1b, however, the ionic channel can exchange ions with the reservoir. Hence, we no longer require the cationic and anionic fluxes to be zero throughout. Instead, only the currents at the channel openings are required to be zero:[21]

$$z(J_+ - J_-)_{x=\pm L/2} = 0 \tag{31}$$

The following equation can be obtained:

$$\left(D_+ \frac{dn_+}{dx} - D_- \frac{dn_-}{dx}\right)_{x=\pm L/2} - 2n_0 \frac{\alpha_+ D_+ - \alpha_- D_-}{T_0} \frac{\Delta T}{L} = 0 \tag{32}$$

With Eq. (20), the quantity $n_m$ has a linear profile. If both reservoirs have the same concentration at $n_0$, then $n_{\pm 1}\left(x = \pm \frac{L}{2}\right) = 0$, and $n_m(x) = \frac{1}{2}(n_{+1} + n_{-1}) = 0$. The concentration gradients of cations and anions therefore have opposite signs $\frac{dn_+}{dx} = -\frac{dn_-}{dx}$.

The boundary conditions for the ionic concentrations are obtained:

$$\left(\frac{dn_+}{dx}\right)_{\pm L/2} = -\left(\frac{dn_-}{dx}\right)_{\pm L/2} = \frac{2n_0}{L}\left(\frac{\alpha_+ D_+ - \alpha_- D_-}{D_+ + D_-}\right) \frac{\Delta T}{T_0} \tag{33}$$

The solution to Eq. (21) for an open 1D system is written as:

$$n_d = 2n_0 \frac{\Delta T}{T_0}\left(\frac{\alpha_+ D_+ - \alpha_- D_-}{D_+ + D_-}\right) \frac{\sinh(\kappa x)}{kL \cosh(\kappa L/2)} \tag{34}$$

By integrating the Poisson equation with the boundary condition $\frac{d\phi}{dx}\left(x = \pm \frac{L}{2}\right) = 0$, we can obtain the electric field and the Seebeck coefficient of a 1D open system:

$$-E = \frac{d\phi}{dx} = \frac{2k_B(\alpha_+ D_+ - \alpha_- D_-)}{ze(D_+ + D_-)} \frac{\Delta T}{L}\left(1 - \frac{\cosh(\kappa x)}{\cosh(\kappa L/2)}\right) \tag{35}$$

$$\tilde{S}_{1D}(\xi) = \frac{2k_B}{ze}\left(\frac{\alpha_+ D_+ - \alpha_- D_-}{D_+ + D_-}\right)\left(1 - \frac{\tanh\xi}{\xi}\right) \tag{36}$$

where the the tilde ~ indicates that the system is open. The bulk limit of the thermopower for an open system is:

$$\tilde{S}_{bulk} = \frac{2k_B}{ze}\left(\frac{\alpha_+ D_+ - \alpha_- D_-}{D_+ + D_-}\right) \tag{37}$$

which is dependent on the diffusivity now. $\tilde{S}_{bulk}$ will also reduce to $S_{bulk}$ once the cation and anion have the same diffusivity: $D_+ = D_- = D$.

### C. Lateral Confinement Effect on Thermopower

This part now focuses on the ionic Seebeck effect in 2D, where the ionic liquid is confined within a narrow channel with a width of $H$, under a temperature gradient over a length of $L$. The channel wall would interact with the electrolyte, forming an electric double layer (EDL). The EDLs induce a lateral electric field along the $y$-axis as shown in Figure 1c-d, in addition to the field due to the ionic Seebeck effect. With this physical picture, the total electric field is written as:

$$\phi(x, y) = \psi(y) + \varphi(x, y) \tag{38}$$

where $\psi(y)$ is the field due to EDL and $\varphi(x, y)$ is the field due to the ionic Seebeck effect. To obtain the perturbative solutions, the concentration profile can be expressed as a perturbation to the Boltzmann distribution:

$$n_\pm(x, y) = e^{\mp\frac{ze\psi}{k_B T_0}}[n_0 + n_{\pm 1}(x, y)] \tag{39}$$

where the extra Boltzmann factor $e^{\mp\frac{ze\psi}{k_B T_0}}$ is due to the EDL field. Similarly, the temperature gradient is set along the $x$ direction, then the ionic flux is obtained by inserting Eq. (39) into Eq. (5):

$$\boldsymbol{J}_\pm = -D_\pm e^{\mp\frac{ze\psi}{k_B T_0}} \left( \nabla n_{\pm 1} + \frac{z_i e n_0}{k_B T_0} \nabla \varphi - \frac{2\alpha_\pm n_0}{k_B T_0} \frac{\Delta T}{L} \hat{\boldsymbol{x}} \right) \qquad (40)$$

where $\hat{\boldsymbol{x}}$ denotes the unit vector along the $x$-direction. The continuity equation $\nabla \cdot \boldsymbol{J}_\pm = 0$ would result in:

$$\nabla^2 n_{\pm 1} \pm \frac{ze n_0}{k_B T_0} \nabla^2 \varphi = 0 \qquad (41)$$

To solve the thermally-induced field $\varphi$, we invoke the Poisson equation:

$$\nabla^2 \phi = \frac{d^2 \psi}{dy^2} + \nabla^2 \varphi = \frac{2ze n_0}{\epsilon} \sinh\left(\frac{ze\psi}{k_B T_0}\right) - \frac{ze}{\epsilon}\left(n_{+1} e^{-\frac{ze\psi}{k_B T_0}} - n_{-1} e^{\frac{ze\psi}{k_B T_0}}\right) \qquad (42)$$

The EDL $\psi(y)$ satisfies the Poisson-Boltzmann equation:

$$\frac{d^2 \psi}{dy^2} = \frac{2ze n_0}{\epsilon} \sinh\left(\frac{ze\psi}{k_B T_0}\right) \qquad (43)$$

Under the low field condition, the Poisson-Boltzmann equation for EDL can be linearized using the Debye-Hückel approximation,[28] and the EDL field can be analytically solved:

$$\psi(y) = \psi_0 \frac{\cosh(\kappa y)}{\cosh(\kappa H/2)} = \frac{\sigma}{\epsilon \kappa} \frac{\cosh(\kappa y)}{\sinh(kH/2)} \qquad (44)$$

where $\psi_0$ and $\sigma$ is the potential and surface charge density at the boundary walls perpendicular to the $y$-direction. Eq. (44) remains a good approximate solution within 5% error compared with the exact numerical solution of Poisson-Boltzmann equation (Eq. (43)), as long as $ze\psi_0/k_B T_0 < 0.5$.

With Eqs. (42-43), the Poisson equation for the field $\varphi$ induced by the Soret effect can be obtained,

$$\nabla^2 \varphi = -\frac{ze}{\epsilon}\left(n_{+1}e^{-\frac{ze\psi}{k_B T_0}} - n_{-1}e^{\frac{ze\psi}{k_B T_0}}\right) \tag{45}$$

Replacing $n_{\pm 1}$ with $n_m = \frac{1}{2}(n_{+1} + n_{-1})$, $n_d = \frac{1}{2}(n_{+1} - n_{-1})$, the Poisson equation for the thermally induced field $\varphi$ can be rewritten as:

$$\nabla^2 \varphi = -\frac{2ze}{\epsilon}\left[n_d \cosh\left(\frac{ze\psi}{k_B T_0}\right) - n_m \sinh\left(\frac{ze\psi}{k_B T_0}\right)\right] \tag{46}$$

Together with Eq. (41), we obtain a pair of coupled linearized PDEs:

$$\nabla^2 n_m = 0 \tag{47}$$

$$\nabla^2 n_d - \kappa^2 \left[n_d \cosh\left(\frac{ze\psi}{k_B T_0}\right) - n_m \sinh\left(\frac{ze\psi}{k_B T_0}\right)\right] = 0 \tag{48}$$

Eqs. (46-48) can now be numerically solved to obtain the unknown variables $\varphi$, $n_d$ and $n_m$. In this work, the Fipy python library is used to simultaneously solve the coupled PDEs through the finite volume method. [29]

In a closed 2D system as shown in Figure 1c, the boundary conditions are:

$$\begin{aligned} J_\pm^x(x = \pm L/2, y) &= 0 \\ J_\pm^y(x, y = \pm H/2) &= 0 \\ \frac{\partial \varphi}{\partial x}(x = \pm L/2, y) &= 0 \\ \frac{\partial \varphi}{\partial y}(x, y = \pm H/2) &= 0 \end{aligned} \tag{49}$$

The boundary conditions for $n_m$ and $n_d$ can therefore be obtained as:

$$\frac{\partial n_m}{\partial x}(x = \pm L/2, y) = \frac{(\alpha_+ + \alpha_-)\Delta T_0}{T_0}\frac{n_0}{L}$$

$$\frac{\partial n_m}{\partial y}(x, y = \pm H/2) = 0$$

$$\frac{\partial n_d}{\partial x}(x = \pm L/2, y) = \frac{(\alpha_+ - \alpha_-)\Delta T_0}{T_0}\frac{n_0}{L} \qquad (50)$$

$$\frac{\partial n_d}{\partial y}(x, y = \pm H/2) = 0$$

The quantity $n_m$ can be analytically solved, which only depends on $x$:

$$n_m(x) = n_0 \frac{(\alpha_+ + \alpha_-)\Delta T}{T_0}\frac{x}{L} \qquad (51)$$

Eq. (48), however, has no simple analytical solution. To proceed analytically, we integrate Eq. (48) along the $y$ direction:

$$\frac{d^2}{dx^2}\int_{-\frac{H}{2}}^{\frac{H}{2}} n_d dy = \kappa^2 \left[\int_{-\frac{H}{2}}^{\frac{H}{2}} n_d \cosh\left(\frac{ze\psi}{k_B T_0}\right) dy - n_m(x)\int_{-\frac{H}{2}}^{\frac{H}{2}} \sinh\left(\frac{ze\psi}{k_B T_0}\right) dy\right] \qquad (52)$$

Further, by assuming that the EDL field is low enough such that the $y$-dependence in $n_d$ is weak, we can approximate:

$$\int_{-\frac{H}{2}}^{\frac{H}{2}} n_d \cosh\left(\frac{ze\psi}{k_B T_0}\right) dy \approx \bar{n}_d \int_{-\frac{H}{2}}^{\frac{H}{2}} \cosh\left(\frac{ze\psi}{k_B T_0}\right) dy \qquad (53)$$

where $\bar{n}_d(x) = \frac{1}{H}\int n_d dy$. An ordinary differential equation for the concentration difference is therefore obtained:

$$\frac{d^2 \bar{n}_d}{dx^2} - \zeta_c \kappa^2 \bar{n}_d = -\kappa^2 \zeta_s n_m(x) \qquad (54)$$

where the factors $\zeta_c$ and $\zeta_s$ are defined as:

$$\zeta_c = \frac{1}{H}\int_{-\frac{H}{2}}^{\frac{H}{2}} \cosh\left(\frac{ze\psi}{k_B T_0}\right) dy$$

$$\zeta_s = \frac{1}{H}\int_{-\frac{H}{2}}^{\frac{H}{2}} \sinh\left(\frac{ze\psi}{k_B T_0}\right) dy \tag{55}$$

The factor $\zeta_c$ is guaranteed to be positive, while the sign of $\zeta_c$ depends on the sign of surface charges. For negative charges on the channel boundary, we have $\zeta_s < 0$, while for positive charges, $\zeta_s > 0$. The solution to Eq. (54) is written as:

$$\bar{n}_d = n_0 \frac{\Delta T}{T_0}\left([(\alpha_+ - \alpha_-) + f_\psi(\alpha_+ + \alpha_-)]\frac{\sinh(\bar{\kappa}x)}{\bar{\kappa}L \cosh\left(\frac{\bar{\kappa}L}{2}\right)} - f_\psi(\alpha_+ + \alpha_-)\frac{x}{L}\right) \tag{56}$$

where $f_\psi = \zeta_s/\zeta_c$, and $\bar{\kappa}$ is the corrected inverse Debyle length due to the electrostatic screening effect by the EDL:

$$\bar{\kappa} = \kappa\sqrt{\zeta_c} = \kappa\left(\int_{-\frac{H}{2}}^{\frac{H}{2}} \cosh\left(\frac{ze\psi(y)}{k_B T_0}\right) dy\right)^{1/2} \tag{57}$$

When there are no surface charges on the lateral boundaries, the EDL field would be zero, and the correction factors $f_\psi = 0$ and $\zeta_c = 1$, thereby Eq. (56) recovers Eq. (26) of the 1D case. Integrating Eq. (46) along the $y$ direction, we can obtain:

$$\frac{d^2\bar{\varphi}}{dx^2} = -\frac{2ze}{\epsilon}[\zeta_c\bar{n}_d(x) - \zeta_s n_m(x)] \tag{58}$$

Similar to the 1D case, integrating Eq. (58), we can obtain the electric field and the Seebeck coefficient as:

$$S_{2D} = \frac{k_B(\alpha_+ - \alpha_-)}{ze}\left[1 - f_\psi\left(\frac{\alpha_+ + \alpha_-}{\alpha_+ - \alpha_-}\right)\right]\left(1 - \frac{\tanh(\bar{\xi})}{\bar{\xi}}\right) \tag{59}$$

where $\bar{\xi} = \bar{\kappa}L/2$. Similar to the 1D case, the Seebeck coefficient is not dependent on diffusivity. When $\bar{\xi} \gg 1$, the lateral boundary effect on the thermopower of a closed system is expressed as:

$$S_{2D}(\bar{\xi} \gg 1) = \frac{k_B(\alpha_+ - \alpha_-)}{ze}\left[1 - f_\psi\left(\frac{\alpha_+ + \alpha_-}{\alpha_+ - \alpha_-}\right)\right] \tag{60}$$

This indicates that the condition for increasing ionic Seebeck coefficient is: $f_\psi(\alpha_+ - \alpha_-)^{-1} < 0$, which is indeed consistent with the physical intuition. For a p-type electrolyte, $\alpha_+ - \alpha_- > 0$, negative surface charges ($f_\psi < 0$) would result in more unipolar charge transport in the electrolyte and hence the increased Seebeck coefficient.

Finally, we turn to solve the equations for the 2D open system where the ionic channel is connected to the two reservoirs as shown in Figure 1d. At the two openings connected to the reservoir, we impose the boundary conditions similar to the 1D case:

$$\begin{aligned}
(J_+^x - J_-^x)_{x=\pm L/2} &= 0 \\
J_\pm^y(x, y = \pm H/2) &= 0 \\
\frac{\partial \varphi}{\partial x}(x = \pm L/2, y) &= 0 \\
\frac{\partial \varphi}{\partial y}(x, y = \pm H/2) &= 0
\end{aligned} \tag{61}$$

With these boundary conditions, we can obtain that $n_m(x) = 0$, and the PDE for $n_d$ is simplified as:

$$\nabla^2 n_d - \kappa^2 \cosh\left(\frac{ze\psi}{k_B T_0}\right) n_d = 0 \tag{62}$$

By following the same derivation by integrating the $y$ coordinates, the concentration profile $n_d$ can be written as:

$$n_d = 2n_0 \frac{\Delta T}{T_0} \left[\frac{\alpha_+ D_+(1-f_\psi) - \alpha_- D_-(1+f_\psi)}{D_+(1-f_\psi) + D_-(1+f_\psi)}\right] \frac{\sinh(\bar{\kappa}x)}{\bar{\kappa}L \cosh(\bar{\kappa}L/2)} \tag{63}$$

and the thermopower can also be derived by solving the $y$-integrated Poisson equation:

$$\tilde{S}_{2D} = \frac{2k_B}{ze}\left[\frac{\alpha_+ D_+(1-f_\psi) - \alpha_- D_-(1+f_\psi)}{D_+(1-f_\psi) + D_-(1+f_\psi)}\right]\left(1 - \frac{\tanh(\bar{\xi})}{\bar{\xi}}\right) \tag{64}$$

With the length approaching infinity, we have therefore derived the relation of liquid electrolyte confined inside a long open channel:

$$\tilde{S}_{2D}(\bar{\xi} \gg 1) = \frac{2k_B}{ze}\left[\frac{\alpha_+ D_+ - \alpha_- D_- - f_\psi(\alpha_+ D_+ + \alpha_- D_-)}{D_+ + D_- - f_\psi(D_+ - D_-)}\right] \tag{65}$$

Similar to the 1D solution, $\tilde{S}_{2D}(\kappa L \gg 1)$ converges to $S_{2D}(\kappa L \gg 1)$ of the closed system when the cations and anions have identical diffusivity $D_+ = D_-$.

### III. Numerical Results and Discussions

To check the validity of our theory, size-dependent thermopower is calculated as a function of dimensionless boundary distances $\kappa H$ and $\kappa L$. As shown in Figure 2a-b, the analytical expressions Eqs. (6) and (65) are compared with the numerical solutions of Eqs. (46-48), for both open and closed systems. Analytical and numerical results are obtained for the p-type electrolyte NaCl solution, with $\alpha_+ = 0.7$, $\alpha_- = 0.1$,[12] and a diffusivity ratio of $D_+/D_- = 0.6394$,[3] at a dimensionless EDL potential of $e\psi_0/k_B T_0 = -1$ where the lateral boundaries are negatively charged. At such high potential, the Poisson equation for EDL (Eq. (43)) can no longer be linearized with the Debye-Hückel approximation. Instead, Eq. (43) is numerically solved using the 4$^{th}$-order collocation algorithm,[30] and the correction factors $\zeta_c$, $\zeta_s$ and $f_\psi$ are then calculated by integrating the EDL potential $\psi(y)$ for evaluating the analytical expression Eqs. (6) and (65). Exellent agreement has been achieved for a wide range of $\kappa H$ and $\kappa L$. It can also been seen from Figure 2a-b that lateral confinement by negatively charged surfaces tends to increase the p-type thermopower, because the overlapping of the EDL breaks the local charge neutrality inside the electrolyte. According to the Poisson equation, the charge density of the EDL is expressed as $\rho(y) = -2zen_0 \sinh\left(\frac{ze\psi}{k_B T_0}\right)$, as such, decreasing the dimensionless distance $\kappa H$ between two negatively charged boundaries ($\psi < 0$) would make the electrolyte more unipolar with the majority charge carrier being the cations. With the same $\kappa H$, the thermopower of closed systems converges to higher value than that of open systems at the limit of $\kappa L \gg 1$. For the open system, the p-type thermopower is suppressed compared with the closed system because the cationic diffusivity of Na$^+$ is smaller than the diffusivity of Cl$^-$. The thermopower in the limit of $\kappa L \gg 1$ is then evaluated for different channel widths $\kappa H$ and EDL potential $e\psi_0/k_B T_0$ as shown in Figure 2c-d. Consistent with the physical intuition, more negative surface potential $\psi_0$ and the EDL overlapping effect at decreasing channel widths $H$ result in increasingly unipolar ionic transport in the electrolyte and an increased p-type thermopower.

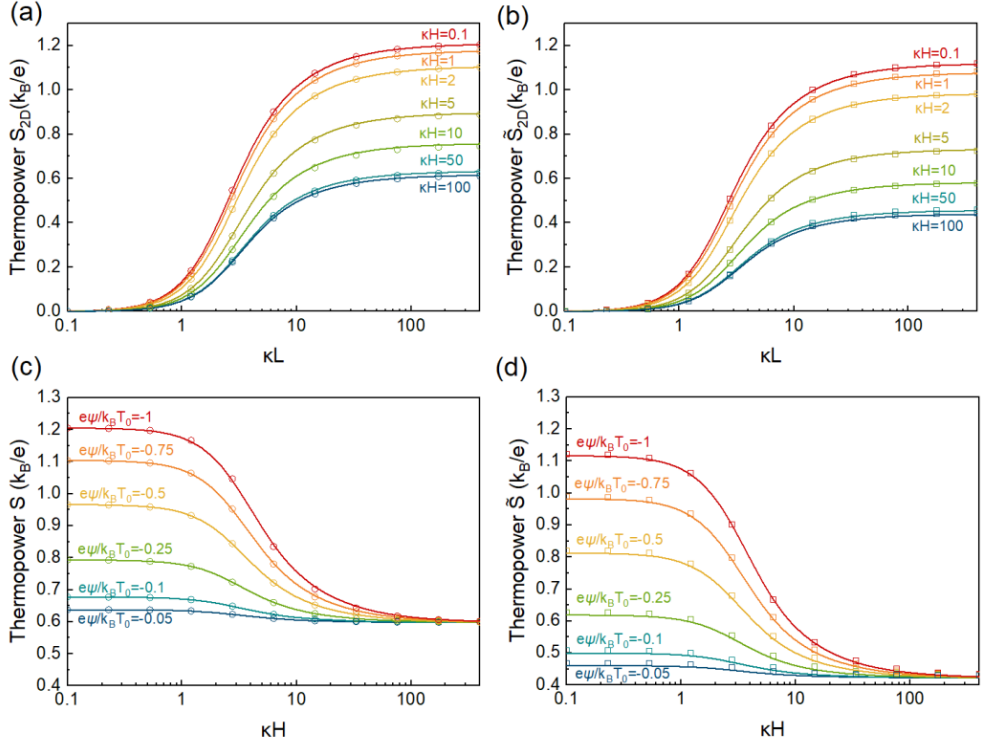

Figure 2. Comparison between the analytical solution of thermopower (solid lines) and numerical solutions (symbols) to Eqs. (46-48) of (a) closed and (b) open systems for various dimensionless boundary distances $\kappa H$ and $\kappa L$. The curves are generated with an EDL potential $e\psi_0/k_B T_0 = -1$. Potential $e\psi_0/k_B T_0$ and width dependence $\kappa H$ of thermopower for (c) closed and (d) open systems.

With two- to three-fold increase in the thermopower of confined NaCl, it would also be interesting to examine the possibilities of further enhancement of the thermopower of polyelectrolytes such as $Na^+PEO^-$ through confinement, whose thermopower is intrinsically high at ~ 11 mV/K.[11] These polyelectrolytes usually have large, weakly mobile polyions and small, mobile counterions, as shown in Figure 3a. A reasonable assumption for the case of $Na^+PEO^-$ is that these polymeric anions are largely immobile, such that $D_+ \gg D_-$. In this case, the measured thermopower is indeed dominated by the $Na^+$ ions.

Under this assumption, the dimensionless Soret coefficient $\alpha_+$ is estimated as large as 127.6. Similarly large Soret coefficient has also been observed in other polyelectrolyte systems,[31] which could possibly be attributed to selective ionic interaction and the thermophoresis of solvent water in the polyelectrolyte.[32, 33] For these polyelectrolytes with largely mismatched diffusivity and Soret coefficients, the confinement induced EDL overlap effect has a negligible effect on thermopower $\tilde{S}$ of the open system. Neglecting $D_-$ and $\alpha_-D_-$ in Eq. (65), the thermopower for the open system is approximately a constant $\tilde{S}_{2D}(\xi \gg 1) \approx 2k_B\alpha_+/e$, and any lateral confinement would have negligible effect on the thermopower. For closed systems, however, there is still great opportunities for further enhancement when confined inside charged nanochannels, as shown in Figure 3b. Using the analytical model, we can further estimate the contribution to thermopower enhancement of Na$^+$PEO$^-$ due to the confinement of nanocellulose channels. With the high porosity of the nanocellulose membrane $p \sim 90\%$,[34] and the channel width 0.6 nm,[22] the specific area per volume can be estimated with $A_e = 4(1-p)/H$.[35] Given the volumetric charge density ~ 0.25 mmol/g and a density of 0.29 g/cm$^3$,[22] the surface charge density $\sigma$ can therefore be estimated to be around -0.02 C/m$^2$ and hence EDL potential $e\psi_0/k_BT$ is close to 0.5. For 0.625 M Na$^+$PEO$^-$ electrolyte with dielectric constant ~42.5,[11] the $\kappa H$ is estimated ~2.1 within 0.6 nm channel inside nanocellulose. From our analytical expression, the maximum thermopower of such confined system is estimated to be ~15 mV/K, smaller than the reported 24 mV/K.[22] Note that the estimation here can be regarded as the upper bound for confinement-enhanced thermopower, since the mobile OH$^-$ ions that tends to suppress the p-type thermopower have been completely neglectied, suggesting that there exist other mechanisms for the increased thermopower observed in experiment. Recent analysis pointed out that the thermpower of complex electrolytes could be proportional to the hopping enthalpy barrier $\Delta H$, contributing to extra heat of transport $Q^*$ responsible for excess thermopower enhancement.[32]

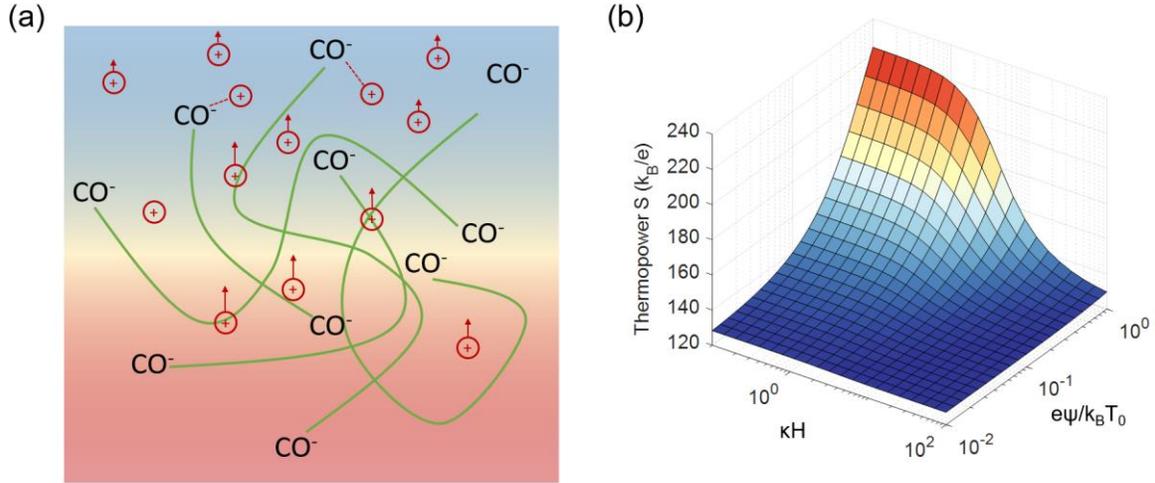

Figure 3. (a) Schematic of Na$^+$PEO$^-$ polyelectrolyte with large polymeric anions and mobile Na$^+$ ions. (b) Estimated thermopower in a closed system with various EDL potentials and channel width.

### IV. Summary and Conclusions

This work presents a theory for the confinement effect on the thermopower of ionic Seebeck effect, by solving the Poisson-Nernst-Planck equations using the first-order perturbation method. Through rigorous solutions to the linearized Posson-Nernst-Planck equations, we have clarified the differences of thermopower for closed and open systems and presented analytical expressions capturing the lateral confinement effect on ionic thermopower. Our analytical expressions showed excellent agreement with numerical solutions in a wide range of channel widths, lengths, and EDL potential at the boundary surfaces. Finally, this article presents insights into the increased thermopower of confined polyelectrolytes with extremely mismatched diffusivities between cations and anions. Our theory showed that lateral confinement can enhance the thermopower of polyelectrolytes only for closed systems, and the increased thermopower could not be explained with mismatched ionic mobilities. This work provides insights into the thermopower enhancement of liquid electrolytes through the nanoscale confinement effect.

**Acknowledgment**

X.Q. acknowledges the startup funding by Huazhong University of Science and Technology. R.Y. acknowledges the National Natural and Science Foundation of China (NSFC) under Grant No. 52036002. The authors declare no conflict of interest.